\documentclass[twocolumn,showpacs,preprintnumbers,amsmath,amssymb]{revtex4}
\usepackage{amsmath,amsfonts,latexsym,amssymb,graphics,epsfig,subfigure,color,makeidx}
\usepackage{lipsum}
\usepackage{float}
\usepackage[dvipsnames]{xcolor}
\usepackage[colorlinks,
            linkcolor=blue,
            anchorcolor=blue,
            citecolor=green
            ]{hyperref}
\usepackage{tikz}
\usepackage{graphicx}
\usetikzlibrary{positioning}
\usetikzlibrary{patterns}

\def\newacronym#1#2#3{\gdef#1{#3 (#2)\gdef#1{#2}}}

\newacronym{\sdsc}{SDSC}{San Diego Supercomputer Center}
\newacronym{\cgp}{CGP}{Center for Gravitational Physics}
\newacronym{\TACC}{TACC}{Texas Advanced Computing Center}
\newacronym{\nr}{NR}{numerical relativity}
\newacronym{\DA}{DA}{data analysis}
\newacronym{\pn}{PN}{post Newtonian}
\newacronym{\PCA}{PC}{Principal Component}
\newacronym{\CBC}{CBC}{compact object coalescences}
\newacronym{\grmhd}{GRMHD}{general relativistic magneto-hydrodynamic}
\newacronym{\mhd}{MHD}{magneto-hydrodynamic}
\newacronym{\ornl}{ORNL}{Oak Ridge National Laboratory}
\newacronym{\lisa}{LISA}{Laser Interferometer Space Antenna}
\newacronym{\ligo}{LIGO}{Laser Interferometer Gravitational Wave Observatory}
\newacronym{\lsc}{LSC}{LIGO Scientific Collaboration}
\newacronym{\lvk}{LVK}{LIGO, Virgo and KAGRA}
\newacronym{\sph}{SPH}{smooth particle hydrodynamics}
\newacronym{\tsi}{TSI}{Terascale Supernova Initiative}
\newacronym{\wmap}{WMAP}{the Wilkinson Microwave Anisotropy Probe}
\newacronym{\cmbr}{CMBR}{cosmic microwave background}
\newacronym{\imbbh}{IMBBH}{intermediate mass binary black hole}
\newacronym{\hpc}{HPC}{High-performance Computing}
\newacronym{\bssn}{BSSN}{Baumgarte Shapiro Shibata Nakamura}
\newacronym{\lvc}{LVC}{LIGO Virgo Collaboration}
\newacronym{\EOB}{EOB}{Effective One Body}
\newacronym{\tat}{TAT}{Theoretische Astrophysik T\"ubingen}
\newacronym{\ninja}{NINJA}{Numerical INJection Analysis}
\newacronym{\nrar}{NRAR}{Numerical Relativity Analytical Relativity}
\newacronym{\CTS}{CTS}{Conformal Thin Sandwich}
\newacronym{\gr}{GR}{General Relativity}
\newacronym{\ml}{ML}{machine learning}
\newacronym{\uta}{UT-Austin}{University of Texas at Austin}
\newacronym{\imr}{IMR}{inspiral-merger-ringdown}
\newacronym{\pe}{PE}{parameter estimation}

\def\cbc#1{compact object coalescence#1 (CBC#1)\gdef\cbc{CBC}}
\def\ahz#1{apparent horizon#1 (AH#1)\gdef\ahz{AH}}
\def\qnm#1{quasi-normal mode#1 (QNM#1)\gdef\qnm{QNM}}
\def\snr#1{signal-to-noise ratio#1 (SNR#1)\gdef\snr{SNR}}
\def\si#1{Senior Investigator#1 (SI#1)\gdef\si{SI}}
\def\emri#1{Extreme Mass-Ratio Inspiral#1 (EMRI#1)\gdef\emri{EMRI}}
\def\imbh#1{intermediate mass black hole#1 (IMBH#1)\gdef\imbh{IMBH}}
\def\imbhb#1{intermediate mass black hole binary#1 (IMBHB#1)\gdef\imbhb{IMBHB}}
\def\smbh#1{supermassive black hole#1 (SMBH#1)\gdef\smbh{SMBH}}
\def\bbh#1{binary black hole#1 (BBH#1)\gdef\bbh{BBH}}
\def\bh#1{black hole#1 (BH#1)\gdef\bh{BH}}
\def\ns#1{neutron star#1 (NS#1)\gdef\ns{NS}}
\def\hmns#1{hypermassive neutron star#1 (HMNS#1)\gdef\hmns{HMNS}}
\def\whd#1{white dwarf#1 (WD#1)\gdef\whd{WD}}
\def\gw#1{gravitational wave#1 (GW#1)\gdef\gw{GW}}
\def\isco#1{innermost stable circular orbit#1 (ISCO#1)\gdef\isco{ISCO}}
\def\EM#1{electromagnetic#1 (EM#1)\gdef\EM{EM}}
\def\pnw#1{post-Newtonian#1 (PN#1)\gdef\pnw{PN}}
\def\eob#1{effective one body#1 (EOB#1)\gdef\eob{EOB}}
\def\eos#1{equation of state#1 (EOS#1)\gdef\eos{EOS}}
\def\grb#1{gamma-ray burst#1 (GRB#1)\gdef\grb{GRB}}
\def\tde#1{tidal disruption event#1 (TDE#1)\gdef\tde{TDE}}
\def\pca#1{Principal Component Analysis#1 (PCA#1)\gdef\pca{PCA}}
\def\bhns#1{black hole - neutron star#1 (BHNS#1)\gdef\bhns{BHNS}}
\def\dns#1{double neutron star#1 (DNS#1)\gdef\dns{DNS}}
\def\jbd#1{Jordan-Brans-Dicke#1 (JBD#1)\gdef\jbd{JBD}}
\def\cbc#1{compact binary coalescence#1 (CBC#1)\gdef\cbc{CBC}}
\def\atg#1{alternative theory of gravity#1 (ATG#1)\gdef\atg{ATG}}

\def\rift#1{Rapid parameter estimation via Iterative FiTing#1 (RIFT#1)\gdef\rift{RIFT}}
  \def\jbd#1{Brans-Dicke#1
  (JBD#1)\gdef\jbd{BD}}
  \def\stg#1{scalar-tensor#1
  (ST#1)\gdef\stg{ST}}

\newcommand{\maya}{\textit{Maya}}

\newcommand{\etk}{\textit{Einstein Toolkit}}

\usepackage{color}
\usepackage{placeins}
\usetikzlibrary{angles,calc,intersections,quotes,arrows.meta}

\begin{document}

\title{Growth of a Black Hole in a Scalar Field Cosmology}

\author{Jake Doherty,
		Miguel Gracia-Linares,
        Pablo Laguna
}
\affiliation
{
Center of Gravitational Physics, Department of Physics, University of Texas at Austin, Austin, TX 78712, U.S.A.
}

\begin{abstract}
We present a numerical relativity study of the accretion properties of a non-spinning black hole in a cosmology driven by a scalar field. The simulations are carried out with a modified moving-puncture gauge condition suitable for cosmological space-times. We considered a scalar field with potential $ V=\lambda \,\varphi^4/4$ and derived the black hole mass growth formula for this scenario using the dynamical horizon framework. As with perturbative studies, we find that the accretion rate $\dot M \propto M^2$ with $M$ the mass of the black hole, and that $\dot M \propto \dot\varphi^2$. We verify that the results of the simulations satisfy the mass growth formula.
Unexpectedly, the dynamics of the scalar field in the neighborhood of the black hole is not significantly different from the behavior of the field far away from the hole. We found situations in which the black hole can growth $\sim 15\%$ of its initial mass before the scalar field reaches the bottom of its potential.
\end{abstract}

\maketitle

\section{Introduction}

 With the current sensitivity of the \gw{} detectors, observations of \bbh{} mergers by the \lvk{} collaboration~\cite{LIGOScientific:2016lio,LIGOScientific:2020tif,LIGOScientific:2019fpa,LIGOScientific:2021sio} show consistency with the assumption that \gr{} is the correct theory of gravity and that the environment in which the systems merge is in a vacuum. As the sensitivity of the detectors improves, 
 \gw{} observations will present us with a unique opportunity to uncover phenomena that could potentially be governed by alternatives to  \gr{} and to the Standard Model of particle physics. The possibility that \gw{} observations could help us decipher the nature of dark matter and dark energy is also not far-fetched. At the very least, \gw{s} will help us to identify the properties of the environments hosting sources of gravitational radiation~\cite{Fedrow2017,Toubiana:AGN_LISA_2021}. Several studies have addressed finding evidence for modified theories of gravity~\cite{LIGOScientific:2021sio, Berti:TGR_2015, Yunes:GWTC1} or physics beyond the Standard Model, e.g., ultralight bosons~\cite{proca_obs, proca_21g}. For massive \bh{s}, in astrophysical scenarios in which gravitational radiation is accompanied with electromagnetic radiation, the presence of gas/dust cannot be ignored; examples are active galatic nuclei~\cite{Graham:AGN_2023, Rowan:AGN_formation, Tagawa:AGN_2020, Ford:AGN_2022, Vajpeyi:AGN_2022, Grobner:AGN_rate, Barry:AGN_2018} or accretion disks~\cite{Khan:accretion, Yunes:accretion_2011, Novikov:1973_accretion}. \bh{} accretion is expected to influence the coalescence and translate into intensity variations in the \gw{s} emitted~\cite{Sberna:AGN_LISA_2022,Vitor:Env_2022,Vitor:Env_2020, Vitor:Env_2014,Vitor:GW_EMRIs} and also affect the final \bh{} characteristics, including the gravitational recoil~\cite{Zhang_2023}.

Several studies have examined the impact of a scalar field environment in the \gw{} emission of compact mergers. These include phenomenological studies about environmental signatures on isolated \bh{s} or EMRIs~\cite{Yunes:accretion_2011,Macedo2013:DMInspiral}, binaries in various modified gravity theories~\cite{Yunes:GW_EMRI,Berti:ST_NoHair,Healy:ST_BBH,Yagi:LISA,Cao:fR_BBH,Hirschmann:EMD} or within axion fields~\cite{Yang:axion}, scalar field dynamics in \bh{} space-times~\cite{Wong:evolution, Alejandro:SFDM_2011} and their phenomena, such as superradiance~\cite{East:Superradiant,East:Superradiant2,Cardoso:KerrST,Zhang:BBH_superradiance}
and scalarisation~\cite{Cardoso:KerrST,Wong:scalarization}, scalar dynamics in the transition from inspiraling \bh{s} to a single perturbed \bh{}~\cite{Bentivegna2008}, scalar radiation from \bbh{} systems~\cite{Healy:2011ef,Berti:2013gfa}, \bbh{} mergers in $f(R)$ theories~\cite{Cao:2013osa} and in dynamical Chern-Simons gravity~\cite{Okounkova:2017yby}, effects of axion-like scalar field environment on \bbh{} mergers~\cite{Yang:2017lpm}, and \bbh{} dynamics in Einstein-Maxwell-dilation theory~\cite{Hirschmann:2017psw} and in scalar Gauss-Bonnet gravity \cite{Witek:2018dmd}. These studies were partially motivated by the possibility of scalar fields to explain the nature of dark matter \cite{Marsh:2015xka} or for scalar-tensor and $f(R)$ theories of gravity~\cite{Wagoner1970,Felice2010,Sotiriou2010}.   

Generally, there are two processes by which a scalar field environment impacts the \bbh{} dynamics. One is accretion. As the \bh{s} grow,  the evolution of the orbital frequency (and therefore that of the emitted \gw{s}) is altered relative to the vacuum case. Also, the mass of the final \bh{} is larger than that of the corresponding \bh{} in vacuum, impacting its ringdown structure and the excitation of ringdown modes. This would be particularly important for massive binaries, for which the ringdown of the final \bh{} dominates the signal inside the detector's band. In this case, environmental effects could be detectable through the ringdown structure~\cite{TGR_IMR,Pang:2018hjb}. The second effect is dissipation or dragging. As the \bh{s} interact with the scalar field environment, they experience dynamical friction~\cite{Valerio:tidal1, Valerio:tidal2}.

Without a potential, a \bbh{} in a homogeneous and initially stationary sea of scalar field will behave exactly as in a vacuum. This can be seen from  $ G_{ab} = 8\pi\,T_{ab}$ with $T_{ab} =\nabla_a\varphi\nabla_b\varphi -g_{ab}\left(\nabla_c\varphi\nabla^c\varphi/2 + V\right)$ and $\nabla^a\nabla_a\varphi = V_{,\varphi}$. If $V=0$, and initially $\varphi=$ constant and $\partial_t \varphi =0$, we get that $G_{ab} = 0$. Thus, a dynamical scalar field is needed to get differences from \bbh{} inspirals and mergers in a vacuum. One can achieve this with an inhomogeneous field, a non-stationary field, or a scalar field potential. Our previous work used an inhomogeneous field (bubble encapsulating the binary) with and without a vanishing potential~\cite{Healy_2012,Zhang_2023}. The reason for using a bubble was so we have an asymptotically flat space-time and also a vacuum in the neighborhood of the binary. 

We propose triggering scalar field dynamics with a potential. To avoid the complexities associated with the zoo of inflationary potential, we will consider a $V=\lambda\,\varphi^4/4$ potential, so the only knob to turn is the parameter $\lambda$. In the present work, we will study a single, non-rotating \bh{}. The first step will be to derive a \bh{} growth formula, and the second will be to check the correctness of the formula with numerical relativity simulations. There exist studies investigating accretion of scalar fields by \bh{s}. Scalar field \bh{} accretion has been investigated under both slow-roll and ultra slow-roll approximations through a perturbative expansion of the Einstein's equations~\cite{Gregory_2018,Croney_2025}. Also, approximate analytical solutions for a dynamical spherically symmetric black hole in the presence of a minimally coupled self-interacting scalar field have been derived~\cite{de_Cesare_2022}. Our study does not make any approximations; we solve the full non-linear set of \gr{} equations. In both ~\cite{Gregory_2018} and ~\cite{de_Cesare_2022}, it was found that the growth rate of the \bh{} was proportional to the square of its mass, $\dot{M} \propto M^2$, bearing a similar resemblance to the standard Bondi accretion rate. On the other hand, in a study investigating a \bh{} in a scalar field cosmology, ~\cite{Almatwi_2024} found a growth rate $\dot{M} \propto M^3$. As in ~\cite{Gregory_2018} and ~\cite{de_Cesare_2022}, we find that $\dot{M} \propto \dot\varphi^2$.

The paper is organized as follows: A summary of the method to construct initial data of a \bh{} in the presence of a dynamical cosmological background driven by a scalar field is presented in Section~\ref{sec:init_data}. Evolution equations for the scalar field and gauge conditions are discussed in Section~\ref{sec:eqn_scalar}. Numerical setup and simulation parameters are given in Section~\ref{scalar-field}. Scalar field evolution results are discussed in  Section~\ref{sec:scalar_dynamics}. Results showing how the dynamics of the system obey the area increase law are presented in Section~\ref{sec:area}. \bh{} mass growth is discussed in Section~\ref{sec:growth}. Conclusions are given in Section~\ref{sec:conclusions}. Quantities are reported in units of the puncture mass $m$ of the \bh{}, with $G = c = 1$. Space-time signature is $(-+++)$. Space-time indices are denoted with Latin letters from the beginning of the alphabet. Spatial tensor indices are denoted with Latin letters from the middle of the alphabet. Also, $()\dot\, \equiv d/dt$, and $()\mathring\, \equiv d/d\tau$ with $t$ and $\tau$ coordinate and proper time, respectively.

\section{Initial Data}\label{sec:init_data}

Under the  3+1 decomposition of the Einstein field equations~\cite{baumgarte_shapiro_2010}, a space-time with a metric $g_{ab}$ is foliated by spacelike hypersurfaces $\Sigma$ with unit time-like normals $n^a$. The initial data consist of the spatial metric $\gamma_{ab}$ intrinsic to $\Sigma$, the extrinsic curvature $K_{ab}$ of $\Sigma$, the energy density $\rho$ and the momentum density $S_a$. These quantities are obtained from
\begin{eqnarray}
    \gamma_{ab} &=& g_{ab}+n_an_b\\
    K_{ab} &=& -\frac{1}{2}\mathcal{L}_{n} \gamma_{ab} = -\gamma_a\,^c\gamma_b\,^d\nabla_cn_d\\
    \rho &=& n^an^bT_{ab}\\ 
    S^a &=& -\gamma^{ab}n^cT_{bc}\,,
\end{eqnarray}
with $\nabla$ covariant differentiation with respect to $g_{ab}$ and $T^{ab}$ the stress-energy tensor.
For our case of a scalar field $\varphi$, the stress energy tensor $T_{ab}$ has the form:
\begin{eqnarray}
T_{ab}&=& \nabla_{a}\varphi \nabla_{b}\varphi-g_{ab} \left( \frac{1}{2}\nabla_{c}\varphi \nabla^{c}\varphi +V \right)\,. \label{eq:Tmunu}
\end{eqnarray}
Therefore,
\begin{eqnarray}
\rho &=& \frac{1}{2}\Pi^2+\frac{1}{2}D^i\varphi D_i\varphi+ V,\label{energydesity}\\
S_i    &=& -\Pi\,D_i\varphi,\label{energyflux}
\end{eqnarray}
where $\Pi \equiv -n^a\nabla_a\varphi$ is the conjugate momentum of $\varphi$ and $V$ its potential. As mentioned before, we set $V = \lambda\, \varphi^{4}/4$.
The operator $D_i$ denotes covariant differentiation associated with $\gamma_{ij}$.

The initial data must satisfy the Hamiltonian and momentum constraint equations. Namely,
\begin{eqnarray}
\mathcal{R}+K^2-K_{ij}K^{ij}&=& 16\pi\rho\label{eq:Ham} \\
D_jK^{ij}-D^iK &=& 8\pi S^i\,,\label{eq:mom}
\end{eqnarray}
respectively. Here, $\mathcal{R}$ is the Ricci scalar in $\Sigma$ and $K$ the trace of $K_{ij}$.

We solve the constraints (\ref{eq:Ham}) and (\ref{eq:mom}) following the York-Lichnerowicz conformal approach \cite{Lichnerowicz1944,York1971,York1972,Cook2000} in which
\begin{eqnarray}
\gamma_{ij}&=&\psi^4\tilde\gamma_{ij}\,,\\
A_{ij} &=&\psi^{-2}\widetilde{A}_{ij}\,,
\end{eqnarray}
where $A_{ij}$ is the traceless part of the extrinsic curvature.
In addition, we introduce $\widetilde \Pi = \psi^{6}\Pi$.
With these  transformations, the Hamiltonian \eqref{eq:Ham} and the momentum \eqref{eq:mom} constraints read respectively:
\begin{eqnarray}
&&\tilde\Delta\psi-\left(\frac{1}{8}\mathcal{\widetilde R}-\pi\widetilde D^i\phi\,\widetilde D_i\phi\right)\psi\nonumber\\
&-&\left(\frac{1}{12}K^2 - 2\,\pi\,V \right)\psi^5 \nonumber\\
&+&\left(\frac{1}{8}\tilde A^{ij}\tilde A_{ij}+\pi\widetilde\Pi^2\right)\psi^{-7} = 0\label{hamconstraint}\\
&&\widetilde D_j\tilde{A}^{ji} - \frac{2}{3}\psi^6\widetilde D^iK= - 8\pi\widetilde\Pi\widetilde D^i\phi\,,\label{momconstraint}
\end{eqnarray}
where  $\mathcal{\tilde R}$ is the Ricci scalar of the conformal space, $\widetilde D_{i}$ denotes covariant differentiation associated with the conformal metric $\tilde\gamma_{ij}$, and $\tilde\Delta \equiv \widetilde D_i\widetilde D^i$. For simplicity, we also set the conformal space to be flat, i.e. $\tilde\gamma_{ij} = \eta_{ij}$. Thus, the Hamiltonian and the momentum constraints become
\begin{eqnarray}
&&\partial^{j}\partial_{j}\psi +\pi\,\psi\, \partial^{j}\phi\, \partial_{j}\phi \nonumber\\
&-&\left(\frac{1}{12}K^2\ - 2\,\pi\,V\right)\psi^5 +\nonumber\\
&+&\left(\frac{1}{8}\tilde A^{ij}\tilde A_{ij}+\pi\widetilde\Pi^2\right)\psi^{-7} 
= 0
\label{hamconstraint2}\\
&&\partial_{j}\tilde{A}^{ji} - \frac{2}{3}\psi^6 \partial^{i}K = - 8\pi\widetilde\Pi \partial^{i}\phi\,. \label{momconstraint2}
\end{eqnarray}

Because our system involves a \bh{}, we will model the hole as a puncture  where the conformal factor has the form 
\begin{equation}
    \psi = 1 + \frac{m}{2\,r} + u\,,
    \label{eq:conformal}
\end{equation}
with $m$ the puncture bare mass parameter. If the \bh{} has momentum or spin, we will be using the Bowen-York solutions~\cite{1980PRDBowen} $\tilde{A}^{ji}$ of the momentum constraint in which  $\partial_{j}\tilde{A}^{ji} =0$. To use these solutions, we must choose $K$, $\varphi$, and $\Pi$ so that the terms involving these quantities in Eq.~(\ref{momconstraint2}) vanish. We accomplish this by setting $\Pi = 0$, $K$ = constant and $\varphi$ = constant, which  implies $\dot\varphi = 0$. With these assumptions Eq.~(\ref{hamconstraint2}) takes the following form
\begin{eqnarray}
&&\partial^{j}\partial_{j}\psi+\frac{1}{8}\tilde A^{ij}\tilde A_{ij}\psi^{-7}\nonumber\\
&-&\left(\frac{1}{12}K^2 - 2\,\pi\,V\right)\psi^5
= 0\,.
\label{hamconstraint3}
\end{eqnarray}
The last term in this equation will diverge at the puncture location because of the form of the conformal factor (\ref{eq:conformal}). To avoid this, we set $K^2 = 24\,\pi\,V$. Notice that in a homogeneous cosmological setup, $K = -3H$, and this equation becomes $H^2 = 8\,\pi\,V/3$, the Friedmann equation for a cosmology driven by a scalar field at initial time when $\dot \varphi = 0$. Therefore, we just need to solve the equation
\begin{eqnarray}
&&\partial^{j}\partial_{j}\psi
+\frac{1}{8}\tilde A^{ij}\tilde A_{ij}\psi^{-7} =0\,
\label{momconstraint3}
\end{eqnarray}
which is the equation commonly solved for vacuum space-times for \bh{} modeled by punctures. 

Since we are focusing on a single, non-spinning \bh{} without linear momentum, $\tilde A^{ij} = 0$, and the solution to Eq.~(\ref{hamconstraint3}) is Eq.~(\ref{eq:conformal}) with $u=0$. In a vacuum space-time, the bare mass parameter would be the mass of the \bh{}. In our case, this is not the case due to the scalar field $\varphi$. The mass $M$ of the \bh{} would be obtained from the area $A$ of its apparent horizon as $M = R/2 = \sqrt{A/16\,\pi}$ with $R$ the areal radius of the \bh{.}

\section{Evolution Equations and Gauge Conditions}\label{sec:eqn_scalar}
We solve the Einstein's equations using the BSSN formulation~\cite{baumgarte_shapiro_2010,Shapiro1999,Shibata1995}. The evolution equation for the scalar field is $\nabla^{a}\nabla_{a}\varphi =V_{,\varphi}$. We decompose this equation into a 3+1 form with the help of the spatial metric $\gamma_{ab}$ and the unit normal $n^a = (1,-\beta^{i})/\alpha$ where $\alpha$ is the lapse function and $\beta^i$ the shift vector. The equation of motion takes the form
\begin{eqnarray}
\partial_{t}\varphi - \beta^{i}\partial_{i}\varphi &=& -\alpha\Pi, \label{eq:phi}\\
\partial_{t}\Pi - \beta^{i}\partial_{i}\Pi &=&-\alpha\,D^iD_i\varphi-D^{i}\alpha D_{i}\varphi\nonumber\\
&+& K\,\Pi+\alpha\,V_{,\varphi} \,.\label{eq:pi}
\end{eqnarray}

For the evolution, we used a modified version moving puncture gauge \cite{Campanelli2006,Baker2006} to evolve $\alpha$ and $\beta^{i}$. The moving puncture gauge commonly used for \bh{s} in vacuum is
\begin{eqnarray}
    (\partial_t-\beta^j\partial_i)\alpha &=& -2\,\alpha\,K \label{eq:mvlapse}\\
    \partial_t\beta^i  &=& \frac{3}{4}B^i\\
    \partial_t B^i&=& \partial_t\bar\Gamma^i-\eta\,B^i
\end{eqnarray}
where $\eta$ is a parameter and $\bar\Gamma^i \equiv -\partial_j\bar \gamma^{ij}$ with $\bar\gamma_{ij} = \chi^4\,\gamma_{ij}$ the conformal metric in the BSSN system of equations~\cite{baumgarte_shapiro_2010,Shapiro1999,Shibata1995}. Far away from the \bh{s}, where asymptotic flatness holds, the moving puncture gauge is consistent with $\alpha = 1$, $\beta^i = 0$, and $\gamma_{ij} = \eta_{ij}$.

In the absence of the \bh{}, our space-time is that of a spatially flat Friedmann-Robertson-Walker cosmology, with a metric given as
\begin{eqnarray}
    ds^2 &=& -dt^2+a^2(t)\,\eta_{ij}\,dx^i\,dx^j\,.\label{eq:frw}
\end{eqnarray}
That is, $\alpha = 1$, $\beta^i = 0$, and $\gamma_{ij} = a^2(t)\,\eta_{ij}$ with the expansion factor obeying the Friedmann equation
\begin{eqnarray}
    H^2 &=& \frac{8\,\pi}{3}\rho = \frac{8\,\pi}{3}\left(\frac{1}{2}\dot \varphi^2+V\right)\,,\label{eq:friedmann}
\end{eqnarray}
and $\varphi$ satisfying
\begin{eqnarray}
\ddot{\varphi} + 3H\dot{\varphi}&=&-V_{,\varphi} \,,
\label{eq:phi_friedmann}
\end{eqnarray}
with $H \equiv \dot a/ a$. Therefore, for our situation of a \bh{} in an expanding cosmology, we must ensure that far away from \bh{} we approach a Friedmann-Robertson-Walker cosmology. 

The moving puncture gauge for the shift vector can be directly applicable to our case since, asymptotically, it has $\beta^i = 0$ as a solution. What needs modification is Eq.~(\ref{eq:mvlapse}) for the lapse. In its current form, this equation does not yield asymptotically $\alpha = 1$ or a constant because $K = -3\,H\ne 0$. To get the correct asymptotic behavior for $\alpha$, we introduce the following modification
\begin{eqnarray}
    (\partial_t-\beta^j\partial_j)\alpha &=& -2\,\alpha\,\left(K +\sqrt{24\,\pi\,\rho}\right) \label{eq:mvlapse2}\,.
\end{eqnarray}
With this, far from the hole, the r.h.s. of Eq.~(\ref{eq:mvlapse2}) will vanish because of the Friedmann equation (\ref{eq:friedmann}). We will refer to this condition as the ``cosmological moving puncture gauge."

\section{Computational Setup and Scalar Field Configurations}\label{scalar-field}

Numerical simulations were performed with the \maya{} code \cite{2003VPPR5ICGoodale, Husa2006, 2012ApJHaas, 2015ApJLEvans,2016PRDClark,2016CQGJani}, our local version of the \etk{} code \cite{EinsteinToolkit:2021_11}. All results are given in units of the puncture mass parameter $m$.
The initial configuration is a \bh{} embedded in a homogeneous scalar field. As evolution proceeds, the scalar field will drive a rapid expansion of space-time, which eventually will prevent us from numerically resolving the \bh{}. We use 12 levels of mesh refinements in a computational box of size $120\,m$ with a grid spacing of $m/840$ for the finest mesh. With this setup, we ensure resolving the \bh{} throughout a few hundred $m$ of evolution; this is enough dynamical evolution to address the questions under consideration. We also impose periodic boundary conditions. Strictly speaking, the presence of the \bh{} breaks initially periodicity. However, the outer boundary is sufficiently far from the \bh{} that its effects are minor, and to a good approximation, periodicity is allowed.

We considered $\bar\lambda  = \lbrace 2,4,6,8,10\rbrace $, where $\lambda \equiv \bar\lambda \times 10^{-4}\,m^{-2}$ and used two types of initial values $\varphi_0$ for the scalar field. In one case, as we vary $\bar\lambda$, we keep $\varphi_0 = 0.9$ constant. For the other type, we vary $\varphi_0$ and keep the initial value of $\bar V_{,\varphi} = \bar\lambda\,\varphi^3_0 = 1.458$ constant. The latter is to ensure that the initial ``force" on the scalar field remains the same as we vary $\bar\lambda$. Table~\ref{tab:parms} shows the values for $\bar\lambda$ and $\varphi_0$ chosen as well as the initial mass $M_0$ of the \bh{.} 

\begin{table}[h!]
\centering
\begin{tabular}{ c|c c c|c c c} 
 \hline
 \hline
 $\bar\lambda$  & $\varphi_{0}$ & $\bar V_{,\varphi}$ & $M_0$  & $\varphi_{0}$ & $\bar V_{,\varphi}$ & $M_0$\\ 
 \hline
 2 & 0.900 & 1.458 & 1.001 & 0.900 & 1.458 & 1.001\\ 
 4 & 0.900 & 2.916 & 1.002 &0.714 & 1.458  & 1.002\\
 6 & 0.900 & 4.374 & 1.003 &0.624 & 1.458  & 1.001\\
 8 & 0.900 & 5.832 & 1.004 &0.567 & 1.458  & 1.001\\
 10 & 0.900 & 7.290 & 1.006 &0.526 & 1.458 & 1.001\\
 \hline
 \hline 
 \end{tabular}
\caption{Initial scalar field configuration parameters.  The initial \bh{} mass $M_0$ is given in units of $m$.}
 \label{tab:parms}
 \end{table}

\begin{figure*}[!htbp]
\includegraphics[angle=0,width=1.0 \textwidth]{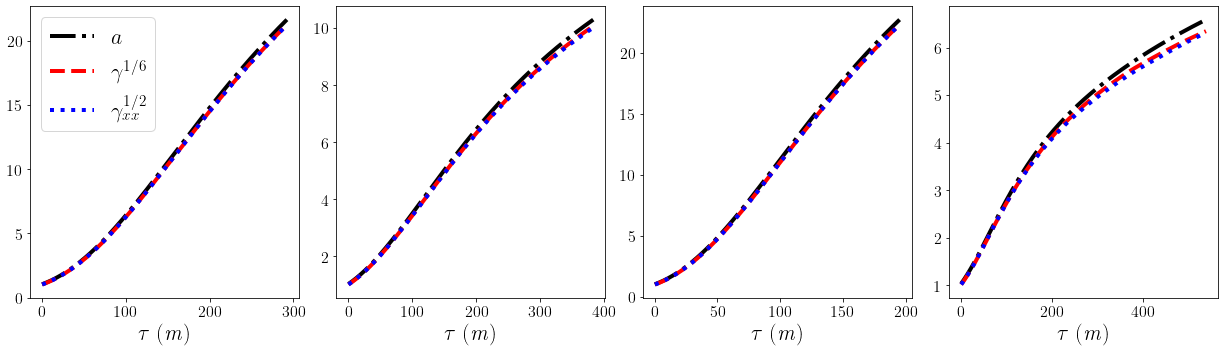}\\
\caption{ $\gamma^{1/6}$, $\gamma_{xx}^{1/2}$, and $a$ as a function of proper time and evaluated at the outer boundary in the $x$-axis. The panels from left to right are for $\lbrace \bar\lambda,\,\varphi_0\rbrace = \lbrace 4, 0.9\rbrace$, $\lbrace 4, 0.714\rbrace$, $\lbrace 10, 0.9\rbrace$, and $\lbrace 10, 0.526\rbrace$, respectively. }
\label{fig:afactor}
\end{figure*}

\begin{figure*}[!htbp]
\includegraphics[angle=0,width=0.45 \textwidth]{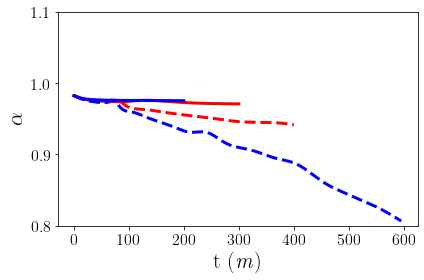}\\
\caption{ Lapse function $\alpha$ as a function of time and evaluated at the outer boundary in the $x$-axis. The lines correspond to  $\lbrace \bar\lambda,\,\varphi_0\rbrace = \lbrace 4, 0.9\rbrace$ (red, solid), $\lbrace 4, 0.714\rbrace$ (red, dashed), $\lbrace 10, 0.9\rbrace$ (blue, solid), and $\lbrace 10, 0.526\rbrace$ (blue, dashed).}
\label{fig:lapse}
\end{figure*}

To demonstrate the ability of the cosmological moving puncture gauge for preserving the standard homogeneous cosmological evolution away from the \bh{,} Figure~\ref{fig:afactor} displays  $\gamma^{1/6}$, $\gamma_{xx}^{1/2}$, and the expansion factor $a$ as a function of proper time at the outer boundary along the $x$ axis. The expansion factor $a$ was obtained by solving the Friedmann equation (\ref{eq:friedmann}). The panels from left to right are for $\lbrace\bar \lambda ,\,\varphi_0\rbrace = \lbrace 4, 0.9\rbrace$, $\lbrace 4, 0.714\rbrace$, $\lbrace 10, 0.9\rbrace$, and $\lbrace 10, 0.526\rbrace$, respectively. The reason for using proper time is because the \bh{} influences the asymptotic value of the lapse function $\alpha$ at the outer boundary, as can be seen in Figure~\ref{fig:lapse}. The fact that $\gamma^{1/6}\approx\gamma_{xx}^{1/2}\approx a$ indicates that far from the \bh{} the space-time behaves as a Friedmann-Robertson-Walker cosmology.

\section{Scalar Field Dynamics}\label{sec:scalar_dynamics}

\begin{figure*}[!htbp]
\includegraphics[angle=0,width=0.8 \textwidth]{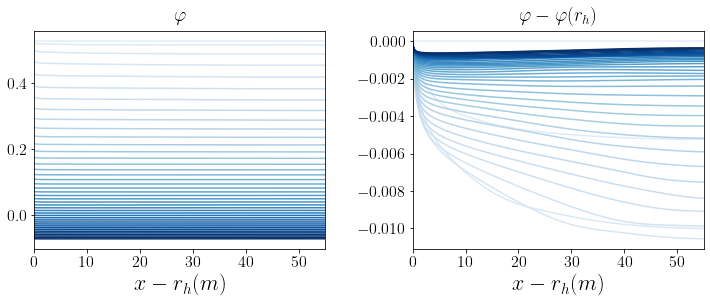}\\
\caption{Scalar field at different times as a function of $x-x_h$ with $x_h$ the horizon location for the case $\lbrace\bar\lambda,\varphi_0\rbrace = \lbrace 10,0.526\rbrace$. Later times correspond to darker tones of blue in the lines. The right panel is identical to the left one, except that the scalar field at the BH horizon is shifted to zero.}
\label{fig:phi_x}
\end{figure*}

\begin{figure*}[!htbp]
\includegraphics[angle=0,width=1.0 \textwidth]{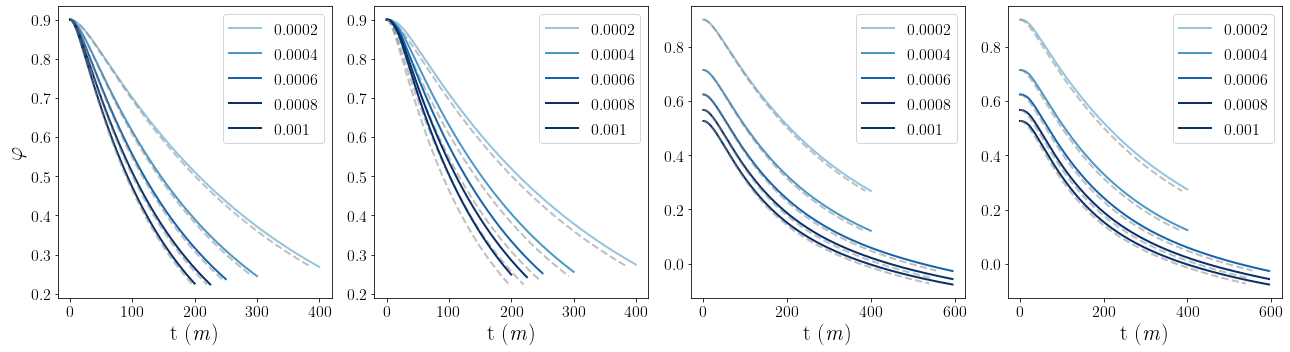}\\
\caption{Evolution of $\varphi$ at the outer boundary in the first and third panels and at the horizon in the second and fourth panels. The first two panels are for the $\varphi_0 = 0.9$ cases and the last two panels for the $\bar V_{,\varphi} = 1.458$ cases. Dashed lines are the values of $\varphi$ from solving Eqs.~(\ref{eq:friedmann}) and (\ref{eq:phi_friedmann}).}
\label{fig:phi_t}
\end{figure*}

\begin{figure*}[!htbp]
\includegraphics[angle=0,width=1.0 \textwidth]{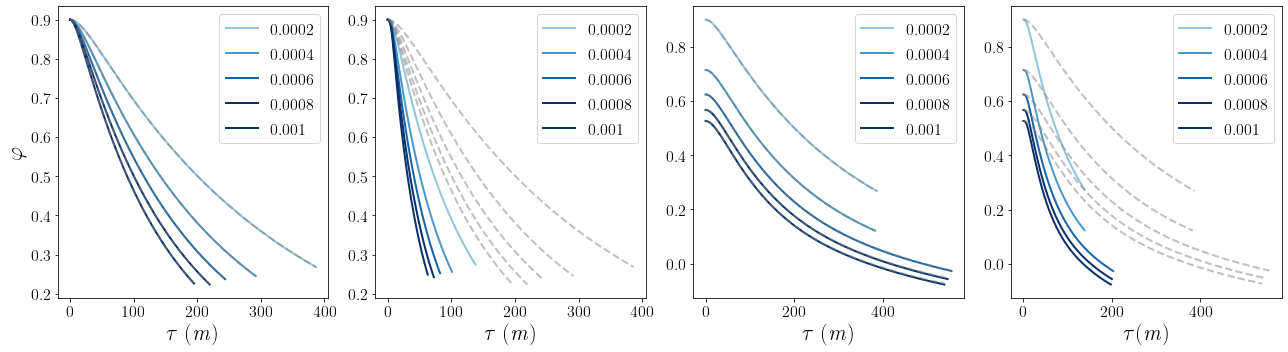}\\
\caption{Same as in Fig.~\ref{fig:phi_t} but as a function of proper time.}
\label{fig:phi_tau}
\end{figure*}

Figure~\ref{fig:phi_x} shows the scalar field $\varphi$ as a function of $x-x_h$ with $x_h$ the horizon location for the case $\lbrace\bar\lambda,\varphi_0\rbrace = \lbrace 10,0.526\rbrace$. Later times correspond to darker tones of blue lines. Surprisingly, as seen in the left panel, the presence of the \bh{} does not significantly modify the scalar field from homogeneity. The differences between the values at the \bh{} horizon and the outer boundary are small, as can be seen in the right panel where we subtract the value at the horizon. As expected, the scalar field decreases in time as it rolls down the potential. However, the decrease at the horizon is slower than far from the hole. This is related to the gravitational redshift near the hole. To help clarify this point, Figure~\ref{fig:phi_t} shows the evolution of $\varphi$ at the outer boundary in the first and third panels and at the horizon in the second and fourth panels. The first two panels are for the $\varphi_0 = 0.9$ cases and the last two panels for the $\bar V_{,\varphi} = 1.458$ cases. Dashed lines are the values of $\varphi$ from solving Eqs.~(\ref{eq:friedmann}) and (\ref{eq:phi_friedmann}).  As expected, the larger the value of $\lambda$, the more rapidly the scalar field evolves toward the bottom of the potential. As noted above, what is remarkable is the similarity of the behavior between the values at the outer boundary and those at the horizon. To stress the difference with the homogeneous cosmological solution, 
in Figure~\ref{fig:phi_tau}, we plot the same results but as a function of proper time. Far from the \bh{}, proper and coordinate time are basically the same since $\alpha \approx 1$.  

\begin{figure*}[!htbp]
\includegraphics[angle=0,width=1.0 \textwidth]{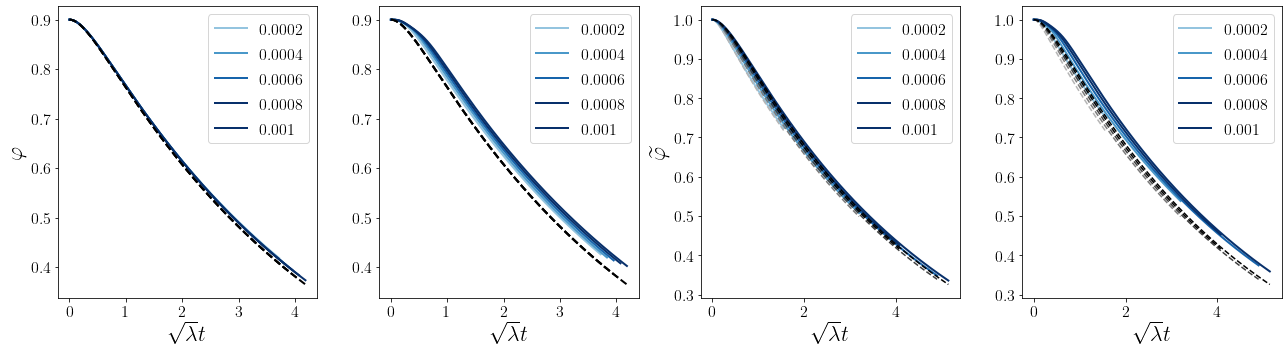}\\
\caption{Same as Fig.~\ref{fig:phi_t} but under the $t \rightarrow \sqrt{\lambda}\,t$. In the last two panels, $\varphi$ has been normalized to its initial value.}
\label{fig:phi_t2}
\end{figure*}

\begin{figure*}[!htbp]
\includegraphics[angle=0,width=1.0 \textwidth]{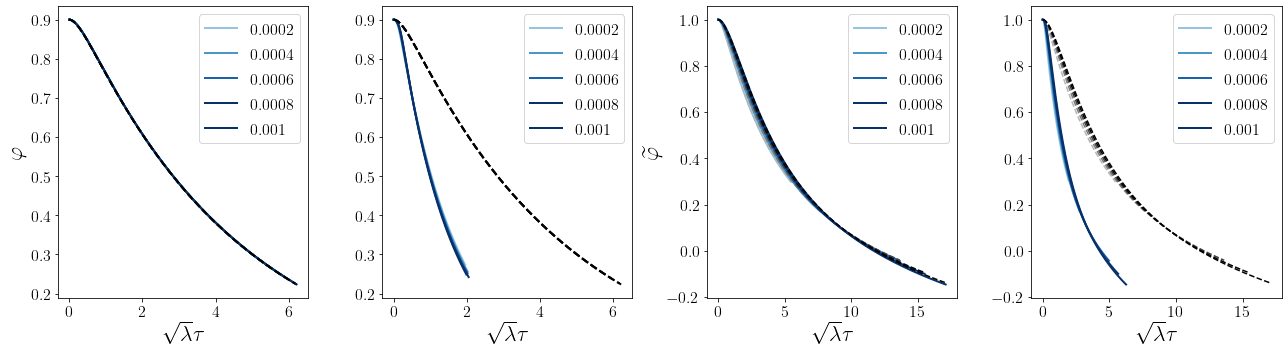}\\
\caption{Same as Fig.~\ref{fig:phi_tau} but under the $t \rightarrow \sqrt{\lambda}\,t$. In the last two panels, $\varphi$ has been normalized to its initial value.}
\label{fig:phi_tau2}
\end{figure*}

For the type of potential we are considering, namely $V = \lambda\,\varphi^4/4$, there is rescaling that effectively eliminates $\lambda$. From the evolution Eqs.~(\ref{eq:phi}) and (\ref{eq:pi}) for the scalar field, this rescaling is  $x^a \rightarrow \sqrt{\lambda}\,x^a$, $\Pi \rightarrow \Pi/\sqrt{\lambda}$, and $K \rightarrow K/\sqrt{\lambda}$. Figures~\ref{fig:phi_t2} and \ref{fig:phi_tau2} are the same as Figures~\ref{fig:phi_t} and \ref{fig:phi_tau}, respectively, under this rescaling.  

\section{Horizon Area Balance Law}\label{sec:area}

According to the first law of \bh{} dynamics, the area of the horizon of a \bh{} in a non-equilibrium situation always increases. The dynamical horizon framework developed by Ashtekar and collaborators provides an expression relating the changes in the area of a \bh{} to the fluxes across its dynamical horizon~\cite{Ashtekar:2004cn}. We will briefly summarize the expression and demonstrate that the outcome of our simulations satisfies such a balance law.

\begin{figure}
 \begin{tikzpicture}
    \draw (3,4.5) arc[start angle=0, end angle=360, x radius=3cm, y radius = 0.3cm] ;
    
    \draw (1.5,2.57) arc[start angle=0, end angle=360, x radius=1.5cm, y radius = 0.2cm] node at (0,2.57) {$S$};


    \draw[-,line width=0.2mm] (-4,2.22) -- (3,2.22);
    \draw[-, line width=0.2mm] (3,2.22) -- (4,2.87);
    \draw[-,line width=0.2mm] (-3,2.87) -- (-2.15,2.87);
    \draw[-,line width=0.2mm] (-3,2.87) -- (-4,2.22) node at (-3,2.45) {$\Sigma$};
    \draw[-, dashed, line width=0.2mm] (-2.15,2.87) -- (2.15,2.87);
    \draw[-, line width=0.2mm] (2.15,2.87) -- (4,2.87);

    \draw[->, thick] (1.5,2.57) -- (1.5,3.57) node[above]{$n^{a}$};
    \draw[->, thick] (1.5,2.57) -- (2.5,2.57) node[right]{$s^{a}$};
    \draw[->, thick, dashed] (1.5,2.57) -- (2.5,3.57) node[above]{$l^{a}_{\Sigma}$};
    \draw[->, thick, dashed] (1.5,2.57) -- (0.5,3.57) node[above]{$k^{a}_{\Sigma}$};

    \draw[->, thick] (-1.5,2.57) -- (-2.5,2.57+0.33) node at (-2.6,3.03) {$r^{a}$};
    \draw[->, thick] (-1.5,2.57) -- (-1.5-0.33,3.57) node[above]{$\tau^{a}$};
    \draw[->, thick, dashed] (-1.5,2.57) -- (-2.5-0.2,3.57+0.2) node[left] {$l^{a}_{H}$};
    \draw[->, thick, dashed] (-1.5,2.57) -- (-0.5-0.2,3.57-0.2) node[right] {$k^{a}_{H}$};

    \draw (3,4.5) arc[start angle=0, end angle=-76, radius=2cm];
    \draw (-3,4.5) arc[start angle=180, end angle=256, radius=2cm];

 \end{tikzpicture}

    \caption{$\Sigma$ is the spacelike hypersurface used in the 3+1 decomposition of the Einstein equations. $S$ is the apparent horizon. $n^a$ is time-like unit normal to $\Sigma$, and $s^a$ is the space-like unit normal to $S$ in $\Sigma$. $l^a_\Sigma$ and $n^a_\Sigma$ are null vectors relative to $n^a$ and $s^a$. $H$ is the dynamical horizon, a space-like hypersurface with time-like unit normal $\tau^a$ and a space-like unit normal $r^a$ normal to $S$ in $H$. $l_H^a$ and $k^a_H$ are null vectors relative to $\tau^a$ and $r_a$.}
         \label{fig:H}
 \end{figure}
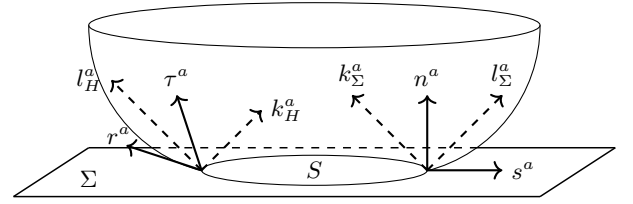

A dynamical horizon $H$ is the world-tube of marginally trapped surfaces or apparent horizons that we will label by $S$. $H$ is foliated by $S$, and $S$ is embedded in the space-like hypersurface $\Sigma$ used in the 3+1 decomposition to solve the Einstein equations. 
With the unit time-like normal $n^a$ to $\Sigma$ and the unit space-like normal $s^a$ to $S$ within $\Sigma$, the following outgoing $l_\Sigma^a$ and ingoing $k^a_\Sigma$ null vectors can be constructed:
\begin{eqnarray}
l_{\Sigma}^{a}  &=& \frac{1}{\sqrt{2}}\left(n^{a}+s^{a}\right) \\ 
k_{\Sigma}^{a}  &=& \frac{1}{\sqrt{2}}\left(n^{a}-s^{a}\right)\,.
\end{eqnarray}
The vectors $n^a$ and $s^a$ in terms of these null vectors are given by
\begin{eqnarray}
\label{eq:nands}
n^{a}  &=& \frac{1}{\sqrt{2}}\left(l^{a}_\Sigma+k^{a}_\Sigma\right) \\
s^{a}  &=& \frac{1}{\sqrt{2}}\left(l^{a}_\Sigma-k^{a}_\Sigma\right)\,.
\end{eqnarray}
Figure~\ref{fig:H} provides a pictorial representation of the setup.

The expansion of $l^a_\Sigma$ and $k^a_\Sigma$  are given by
\begin{eqnarray}
    \Theta_{(l_\Sigma)} &=& h^{ab}\nabla_al^\Sigma_b\\
     \Theta_{(k_\Sigma)} &=& h^{ab}\nabla_ak^\Sigma_b\,,
\end{eqnarray}
respectively, with
\begin{equation}
\label{eq:metric_sigma}
    h_{ab} = g_{ab}+k^\Sigma_al^\Sigma_b+l^\Sigma_ak^\Sigma_b = g_{ab}+n_an_b-s_as_b 
\end{equation}
the 2-dimensional metric in $S$. A trapped surface is one in which $\Theta_{(l_\Sigma)} = 0$ and $\Theta_{(k_\Sigma)} < 0$. The outermost of these surfaces is the apparent horizon. 

The dynamical horizon $H$ is also a space-like hypersurface. It has a unit time-like normal $\tau^a$ and unit space-like normal $r^a$ to $S$ within $H$ (see Fig.~\ref{fig:H}). With $\tau^a$ and $r^a$, the following null vectors can be constructed:
\begin{eqnarray}
l_{H}^{a}  &=& \sqrt{\frac{C}{2}}\left(\tau^{a}+r^{a}\right) \\
k_{H}^{a}  &=& \frac{1}{\sqrt{2\,C}}\left(\tau^{a}-r^{a}\right)\,,
\end{eqnarray}
with $C$ a scalar field fixing the normalization of the null vectors.
Equivalently,
\begin{eqnarray}
\tau^{a}  &=& \frac{1}{\sqrt{2\,C}}\left(l^{a}_H+C\,k^{a}_H\right) \\
r^{a}  &=& \frac{1}{\sqrt{2\,C}}\left(l^{a}_H-C\,k^{a}_H\right)\,.
\label{eq:tauandr}
\end{eqnarray}
Similarly, the expansion of $l^a_H$ and $k^a_H$  are given by
\begin{eqnarray}
    \Theta_{(l_H)} &=& h^{ab}\nabla_al^H_b\\
     \Theta_{(k_H)} &=& h^{ab}\nabla_ak^H_b\,,
\end{eqnarray}
respectively, and the 2-metric (\ref{eq:metric_sigma}) in $S$ takes the form 
\begin{equation}
\label{eq:metric_H}
    h_{ab} = g_{ab}+k^H_al^H_b+l^H_ak^H_b = g_{ab}+\tau_a\tau_b-r_ar_b \,.
\end{equation}
Here again, for an apparent horizon $\Theta_{(l_H)} = 0$ and $\Theta_{(k_H)} < 0$.

Data from numerical relativity simulations are computed in $\Sigma$. On the other hand, the area increase law from the dynamical horizon framework is given in terms of quantities in $H$. We thus need to translate from one hypersurface to the other. The null vectors in $\Sigma$ and $H$ are related to each other via a boost transformation, which in the case of null vectors translates into a multiplicative factor $f$ such that
\begin{eqnarray}
l_{H}^{a}  =  f\,l_\Sigma^{a}\quad\text{and}\quad k_{H}^{a}  = \frac{1}{f}\, k_\Sigma^{a}\,. \label{null_H}
\end{eqnarray}
Notice that with these transformations, 
\begin{eqnarray}
\Theta_{(l_H)}  = f\,\Theta_{(l_\Sigma)}\quad\text{and}\quad
\Theta_{(k_H)}  = \frac{1}{f}\,\Theta_{(k_\Sigma)}\,,
\end{eqnarray}
and
\begin{eqnarray}
r^{a}  &=& \frac{1}{\sqrt{2\,C}} \left( f\,l^{a}_{\Sigma}-\frac{C}{f}\,k^{a}_{\Sigma} \right) \label{eq:r2}\\
\tau^{a}  &=& \frac{1}{\sqrt{2\,C}} \left( f\,l^{a}_{\Sigma}+\frac{C}{f}\,k^{a}_{\Sigma} \right)\,. 
\label{eq:tau2}\label{eq:tau}
\end{eqnarray}

Consider a horizon evolution vector field 
\begin{eqnarray}
    \mathcal{U}^a  &=& U\,r^a \\ \nonumber
    &=& \frac{U}{\sqrt{2\,C}}(l^a_H-C\,k^a_H)\\ \nonumber
    &=& \frac{U}{\sqrt{2\,C}}\left(f\,l^a_\Sigma-\frac{C}{f}\,k^a_\Sigma\right)\,.
    \label{eq:U}
\end{eqnarray} 
The vector field $\mathcal{U}^a$ is tangent to $H$, orthogonal to $S$, and generates a flow preserving the foliation in $H$. That is, $\mathcal{L}_{\mathcal{U}} \Theta_{(l_H)}=0$, which implies that 
 \begin{eqnarray}
     \mathcal{L}_{l_H} \Theta_{(l_H)}-C\,\mathcal{L}_{k_H} \Theta_{(l_H)} &=&0\\
    f^2 \mathcal{L}_{l_\Sigma} \Theta_{(l_\Sigma)}-C\,\mathcal{L}_{k_\Sigma} \Theta_{(l_\Sigma)} &=&0 \,.
 \end{eqnarray} 
Thus,
\begin{eqnarray}
\frac{C}{f^2}  &=& \frac{\mathcal{L}_{l_\Sigma} \Theta_{(l_\Sigma)}}{\mathcal{L}_{k_\Sigma} \Theta_{(l_\Sigma)}} \,.
\label{eq:C}
\end{eqnarray}

From the Raychaudhuri equation,
\begin{eqnarray}
\mathcal{L}_{l_\Sigma} \Theta_{(l_\Sigma)} &=& -\sigma^2-R_{ab}l^a_\Sigma l^b_\Sigma\nonumber\\
&=&- 8\pi\,T_{ab}l^a_\Sigma l^b_\Sigma\,,\label{eq:a1}
\end{eqnarray}
and from $G_{ab}l^a_\Sigma k^b_\Sigma = 8\,\pi\,T_{ab}l^a_\Sigma k^b_\Sigma$, 
\begin{eqnarray}
\mathcal{L}_{k_\Sigma} \Theta_{(l_\Sigma)} &=& -\frac{\cal{\bar R}}{2} + 8\,\pi\, T_{ab}l^a_\Sigma k^b_\Sigma \nonumber\\
&=& - \frac{1}{R^2} + 8\,\pi\,T_{ab}l^a_\Sigma k^b_\Sigma\,,\label{eq:a2}
\end{eqnarray}
where $\mathcal{\bar R} = 2/R^2$ is the scalar curvature of the apparent horizon. Thus, substitution of Eqs.~(\ref{eq:a1}) and (\ref{eq:a2}) into Eq.~(\ref{eq:C}) yields
\begin{eqnarray}
\frac{C}{f^2} &=& \frac{8\,\pi\,R^2\,T_{ab}l^{a}_{\Sigma}l^{b}_{\Sigma}}{1-8\,\pi\,R^2\,T_{ab}l^{a}_{\Sigma}k^{b}_{\Sigma}}\,.
\label{eq:C2}
\end{eqnarray}
From   Eq.~(\ref{eq:Tmunu}) for the stress-energy tensor, we have that
\begin{eqnarray}
    T_{ab}l^a_\Sigma l^b_\Sigma &=&   (l^{a}_{\Sigma}\nabla_{a} \varphi)^2 \nonumber\\
    &=& \frac{1}{2}[( n^a+s^a)\nabla_a\varphi ]^{2}\nonumber\\
    &=&\frac{1}{2}(-\Pi+s^i\partial_i\varphi)^2\nonumber\\
        &=&\frac{1}{2}\left[\mathring\varphi-\left(\frac{\beta^i}{\alpha}-s^i\right)\partial_i\varphi\right]^2\nonumber\\
        &=&\frac{1}{2}(\varphi')^2\label{eq:Tll}\,,
\end{eqnarray}
where we have introduced the following definition:
\begin{equation}
  \varphi'  = \mathring\varphi-\left(\frac{\beta^i}{\alpha}-s^i\right)\partial_i\varphi\,.
  \label{eq:def_prime}
\end{equation}
Similarly,
\begin{eqnarray}
    T_{ab}l^a_\Sigma k^b_\Sigma &=&   l^{a}_{\Sigma}\nabla_{a} \varphi k^{b}_{\Sigma}\nabla_{b} \varphi + \frac{1}{2}\nabla_c\varphi\nabla^c\varphi + V  \nonumber\\
    &=& \frac{1}{2} \nabla_{a} \varphi \nabla_{b} \varphi (g^{ab}+l^a_\Sigma k^b_\Sigma + k^a_\Sigma l^b_\Sigma)+V\nonumber\\
    &=& \frac{1}{2}h^{ab}\nabla_{a} \varphi \nabla_{b}\phi + V = V\,,\label{eq:Tlk}
\end{eqnarray}
where in the last step, we used that the projection of the gradients of $\phi$ on $S$ vanishes because of the spherical symmetry.
With Eqs.~(\ref{eq:Tll}) and (\ref{eq:Tlk}), Eq.~(\ref{eq:C2}) becomes
\begin{eqnarray}
\frac{C}{f^2} &=& \frac{4\,\pi\,R^2\,\left(\varphi'\right)^2}{1-8\,\pi\,R^2\,V}\,.\label{eq:C3}
\end{eqnarray}

The dynamical horizon framework~\cite{Ashtekar:2004cn} tells us that the area increase law reads
\begin{eqnarray}
    \frac{1}{2}\int dR   &=& \mathcal{F}^{(\text{m})}+\mathcal{F}^{(\text{g})}\,
    \label{eq:area}.
\end{eqnarray}
The irreducible mass of the \bh{} is obtained from $M= R/2 = \sqrt{A/16\,\pi}$. In Eq.(\ref{eq:area}),
\begin{eqnarray}
\mathcal{F}^{(\text{m})}  = \sqrt{2} \int_{\Delta H} T_{ab} \tau^{a}\xi_{R}^{b}d^{3}\mathcal{V}
\label{eq:fm}
\end{eqnarray}
is the flux of matter-energy associated with $\xi^a_R$, and
\begin{eqnarray}
\mathcal{F}^{(\text{g})}  = \frac{1}{16\,\pi}\int_{\Delta H}N_R(\sigma^2+2\zeta^2)d^3\mathcal{V}
\label{eq:fg}
\end{eqnarray}
is the energy carried by gravitational radiation. The vector field $\xi^a_R$ is given by $\xi^a_R = N_R\,\,l^a_H$ with $N_R = |\partial R|_H$ a lapse function. $\sigma^2 = \sigma^{ab}\sigma_{ab}$ is the shear and $\zeta^2 = \zeta^a\zeta_a$ with 
$\zeta^a =  h^{ab} r^c\nabla_cl^H_b$. Because we are dealing with a spherically symmetric case, there is no gravitational radiation emitted, and thus $\mathcal{F}^{(\text{g})}=0$. Also, using
$d^3\mathcal{V} = N_R^{-1}dR\,d^2\mathcal{V}$, one can rewrite $\mathcal{F}^{(\text{m})}$ as 
\begin{eqnarray}
    \mathcal{F}^{(\text{m})}   &=& 
    \sqrt{2}\int dR \oint  T_{ab} \tau^{a}l_{H}^{b}d^{2}\mathcal{V}\nonumber\\
    &=& \frac{1}{\sqrt{2}}\int 8\,\pi\,R^2\,T_{ab} \tau^{a}\,l_{H}^{b}\,dR\,,
    \label{eq:area2}
\end{eqnarray}
where we used in the last step that our system is spherically symmetric.
Next, we use $\tau^a$ from Eq.~(\ref{eq:tau}) and $l^a_H$ from Eq.~(\ref{null_H}) and rewrite $\mathcal{F}^{(\text{m})}$ as
\begin{eqnarray}
\mathcal{F}^{(\text{m})} 
&=& \frac{1}{2} \int 8\,\pi\,R^2\,T_{ab} \frac{1}{\sqrt{C}} \left(f\, l^{a}_{\Sigma}+\frac{C}{f}\,k^{a}_{\Sigma} \right) f\,l^b_{\Sigma}\,dR \nonumber \\
&=&\frac{1}{2}\int \sqrt{C}\left( 8\,\pi\,R^2\,T_{ab}l^{a}_{\Sigma}l^b_{\Sigma}\,f^2/C\right. \nonumber\\
&+& \left. 8\,\pi\,R^2\,T_{ab}l^{a}_{\Sigma}k^b_{\Sigma} \right) dR \,.
\label{eq:fm2}
\end{eqnarray}
Finally, if we use Eq.~(\ref{eq:C2}) for $C/f^2$, we get that
\begin{eqnarray}
    \mathcal{F}^{(\text{m})} &=& \frac{1}{2}\int \sqrt{C}\,dR\,.
\end{eqnarray}
Therefore, a comparison with the area law Eq.~(\ref{eq:area}) yields that the normalization field is $C=1$.

\section{Black Hole Mass Growth}
\label{sec:growth}

Next, we derive the expression for the rate of area increase starting from
\begin{eqnarray}
    \mathcal{L}_{\mathcal{U}} \ln{\sqrt{h}} &=& h^{ab}\nabla_a\mathcal{U}_b\,,
\end{eqnarray}
where $h$ is the determinant of the metric $h_{ab}$ in $S$.
From Eq.(\ref{eq:U}), we have that
\begin{eqnarray}
\mathcal{L}_{\mathcal{U}} \ln{\sqrt{h}} &=&\frac{U}{\sqrt{2}}\left(f\, h^{ab}\nabla_a l^\Sigma_b- \frac{1}{f} h^{ab}\nabla_a k^\Sigma_b\right) \nonumber\\
 &=& \frac{U}{\sqrt{2}}\left( f\Theta_{(l_\Sigma)} -\frac{1}{f} \,\Theta_{(k_\Sigma)} \right) \nonumber\\
 &=&-\frac{U}{\sqrt{2}\,f}\,\Theta_{(k_\Sigma)}\,.
 \label{eq:lie}
\end{eqnarray}
Thus,
\begin{eqnarray}
  \mathcal{L}_{\mathcal{U}}\sqrt{h} &=& -\frac{U}{\sqrt{2}\,f}\,\Theta_{(k_\Sigma)}\sqrt{h}\,.
\end{eqnarray}
The area element in $S$ is $\sqrt{h}\,d^2\Omega$. Therefore,
\begin{eqnarray}
  \mathcal{L}_{\mathcal{U}}\int_S\sqrt{h}\,d^2\Omega &=& -\int_S\frac{U}{\sqrt{2}\,f}\,\Theta_{(k_\Sigma)}\sqrt{h}\,d^2\Omega\nonumber\\
   \mathcal{L}_{\mathcal{U}} R^2 &=& -R^2\,\frac{U}{\sqrt{2}\,f}\,\Theta_{(k_\Sigma)}\nonumber\\
    \mathcal{L}_{\mathcal{U}} R &=& -\frac{R}{\sqrt{2}}\,\frac{U}{2\,f}\,\Theta_{(k_\Sigma)}\,,
    \label{eq:dRdV}
\end{eqnarray}
where in the second to last equation, we used the spherical symmetry of our problem, and the expression is evaluated at $S$. 

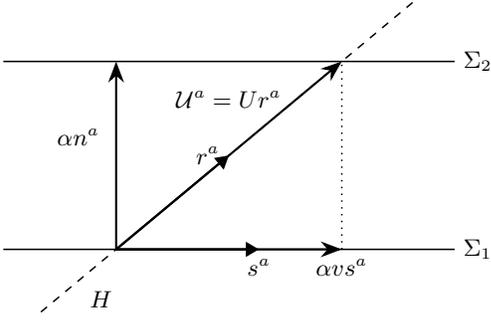
\begin{figure}
 \begin{tikzpicture}

    \draw[-,line width=0.2mm] (-3,2.5) -- (3,2.5) node[right] {$\Sigma_{2}$};

    \draw[-,line width=0.2mm] (-3,0) -- (3,0) node[right] {$\Sigma_{1}$};
    
    \draw[->, thick, >={Stealth[length=8pt]}] (-1.5,0) -- (-1.5,2.5) node at (-2,1.5){$\alpha n^{a}$};

    \draw[->, thick, >={Stealth[length=8pt]}] (-1.5,0) -- (1.5,0) node[below]{$\alpha v s^{a}$};

    \draw[-,line width=0.2mm, dotted] (1.5,0) -- (1.5,2.5);

    \draw[->, thick, >={Stealth[length=8pt]}] (-1.5,0) -- (1.5,2.5) node at (0,2) {$ \mathcal{U}^{a}=Ur^{a}$};

    \draw[->, thick, >={Triangle[length=5pt]}] (-1.5,0) -- (0,1.25) node[left] {$r^{a}$};

    \draw[->, thick, >={Triangle[length=5pt]}] (-1.5,0) -- (0.4,0) node[below] {$s^{a}$};

    \draw[-,line width=0.2mm, dashed] (-2.5,-5/6) -- (2.5,10/3) node at (-1.7, -4/6) {$H$};

 \end{tikzpicture}

\caption{Pictorial representation of the vector $\mathcal{U}^a$ relative to the hypersurfaces $\Sigma$ and the dynamical horizon $H$.}
         \label{fig:H2}
\end{figure}

It is always possible to write $\mathcal{U}^a= \alpha\,(n^a + v\,s^a)$
where $v$ is the coordinate velocity of the apparent horizon in $\Sigma$ (see Fig.~\ref{fig:H2}).
On the other hand, from Eq.~(\ref{eq:U}), we have that
\begin{eqnarray}
    \mathcal{U}^a &=& \frac{U}{\sqrt{2}}(f\,l^a_\Sigma-\frac{1}{f}k^a_\Sigma)\nonumber\\
   &=& \frac{U}{2}\left[f\,(n^a+s^a) - \frac{1}{f} (n^a-s^a) \right]\nonumber \\
   &=& \frac{U\,f}{2}\left[\left(1 - \frac{1}{f^2}\right)\,n^a + \left(1 + \frac{1}{f^2}\right)\,s^a\right]\,.
    \label{eq:vfield}
\end{eqnarray}
Comparing the above expression with $\mathcal{U}^a= \alpha\,(n^a + v\,s^a)$ yields 
\begin{eqnarray}
    \alpha &=& \frac{U\,f}{2}\left(1-\frac{1}{f^2}\right)\label{eq:alpha}\\
    \alpha\,v &=& \frac{U\,f}{2}\left(1+\frac{1}{f^2}\right)\,.\label{eq:valpha}
\end{eqnarray}
With Eq.~(\ref{eq:alpha}), one can eliminate $U$ from Eq.~(\ref{eq:dRdV}) and get
\begin{eqnarray}
     \mathcal{L}_{\mathcal{U}}R &=& -\frac{R}{\sqrt{2}}\alpha\,\Theta_{(k_\Sigma)}\frac{1}{f^2}\,\left(1-\frac{1}{f^2}\right)^{-1}\,.
  \label{eq:dRdV2}
\end{eqnarray}
Substituting Eq.~(\ref{eq:C2}) into this expression, we get that
\begin{eqnarray}
  \mathcal{L}_{\mathcal{U}} R &=&  -\frac{R}{\sqrt{2}}\alpha\,\Theta_{(k_\Sigma)}\times\nonumber\\
  &&\left(\frac{8\,\pi\,R^2\,T_{ab}l^{a}_{\Sigma}l^{b}_{\Sigma}}{1-8\,\pi\,R^2\,T_{ab}l^{a}_{\Sigma}l^{b}_{\Sigma}-8\,\pi\,R^2\,T_{ab}l^{a}_{\Sigma}k^{b}_{\Sigma}}\right)\,,
\end{eqnarray}
or for our particular case
\begin{eqnarray}
 \mathcal{L}_{\mathcal{U}} R &=&  -\frac{\sqrt{8}\,\pi\,\alpha\,\Theta_{(k_\Sigma)}\,R^3\,\varphi'^2}{1-8\,\pi\,R^2\,\left[\frac{1}{2}(\varphi')^2+V\right]}\,.\label{eq:rdot2}
\end{eqnarray}
With $t^a = \alpha\,n^a +\beta^a$, the vector field $\mathcal{U}^a = \alpha(n^a+v\,s^a)$ can be rewritten as $\mathcal{U}^a = t^a - \beta^a+\alpha\,v\,s^a$. Thus,
\begin{eqnarray}
    \mathcal{L}_{\mathcal{U}} R &=& \mathcal{U}^a\nabla_a R\nonumber\\
    &=& t^a\,\nabla_a R + (\alpha\,v\,s^a-\beta^a)\,\nabla_a R\nonumber \\
    &=&  \dot R\,,
\end{eqnarray}
where we have used that the \bh{} areal radius is independent of the spatial coordinates in $\Sigma$.
Finally, 
\begin{eqnarray}
\dot R &=&-\frac{\sqrt{8}\,\pi\,\alpha\,\Theta_{(k_\Sigma)}\,R^3\,\varphi'^2}{1-8\,\pi\,R^2\,\left[\frac{1}{2}\varphi'^2+V\right]}\,,\label{eq:rdot3}
\end{eqnarray}
or in terms of the \bh{} mass
\begin{eqnarray}
\dot M &=&-\frac{8\,\sqrt{2}\,\pi\,\alpha\,\Theta_{(k_\Sigma)}\,M^3\,\varphi'^2}{1-32\,\pi\,M^2\,\left[\frac{1}{2}\varphi'^2+V\right]}\,.\label{eq:mdot}
\end{eqnarray}

\begin{figure*}[!htbp]
\includegraphics[angle=0,width=0.9 \textwidth]{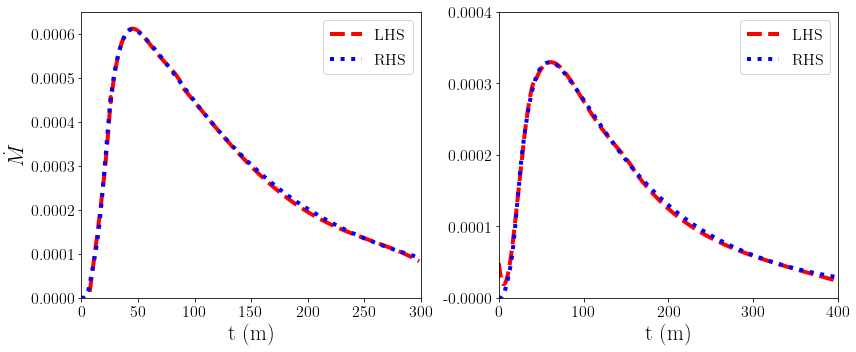}\\
\caption{\bh{} mass growth rate as a function of time; $\lbrace\bar\lambda,\varphi_0\rbrace = \lbrace 4, 0.714\rbrace $ left panel and $\lbrace 4, 0.9\rbrace $ right panel.
The red dashed line shows the growth rate obtained from the area of the apparent horizon in the simulation, and the blue dotted line was obtained by directly substituting numerical data into Eq.~(\ref{eq:mdot}). }
\label{Fig8}
\end{figure*}

\begin{figure*}[!htbp]
\includegraphics[angle=0,width=0.9 \textwidth]{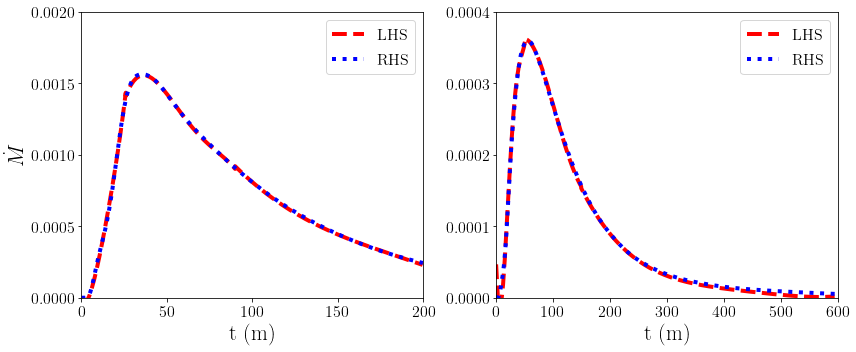}\\
\caption{\bh{} mass growth rate as a function of time; $\lbrace\bar\lambda,\varphi_0\rbrace = \lbrace 8, 0.526\rbrace $ left panel and $\lbrace 8, 0.9\rbrace $ right panel.
The red dashed line shows the growth rate obtained from the area of the apparent horizon in the simulation, and the blue dotted line was obtained by directly substituting numerical data into Eq.~(\ref{eq:mdot}). }
\label{Fig11}
\end{figure*}

As in both ~\cite{Gregory_2018} and ~\cite{de_Cesare_2022}, we find that $\dot{M} \propto \dot\varphi^2$ since $\varphi' \propto \dot\varphi$. These studies also find that $\dot{M} \propto M^2$, similar to the standard Bondi accretion rate. Formula (\ref{eq:mdot}), on the other hand, seems to imply that $\dot{M} \propto M^3$, as in the study in Ref.~\cite{Almatwi_2024}. However, we found that, as with the Schwarzschik-Vaidya metrics~\cite{Ashtekar:2004cn},
$\Theta_{(k_\Sigma)} = - \epsilon/M$, with $\epsilon \approx \mathcal{O}(1)$; thus, $\dot{M} \propto M^2$.

To check the correctness of the mass growth rate as given by Eq.~(\ref{eq:mdot}), we show in Figures~\ref{Fig8} and \ref{Fig11} with red dashed lines $\dot M$ calculated from the area of the apparent horizon and as a blue dotted line $\dot M$ calculated from substituting numerical data into the r.h.s. of Eq.~(\ref{eq:mdot}), demonstrating a clear agreement. 

\begin{figure*}[!htbp]
\includegraphics[angle=0,width=0.9 \textwidth]{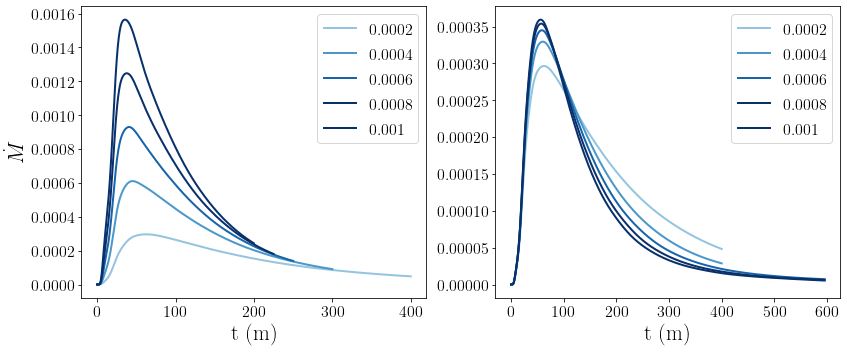}\\
\caption{\bh{} mass growth rate as a function of time. The left panel is for $\varphi_0 = 0.9$ cases, and the right panel is for the  $\bar V_{,\varphi} = 1.458$ cases. Plots were created via Eq.~\ref{eq:mdot}.}
\label{fig:mdot1}
\end{figure*}

\begin{figure*}[!htbp]
\includegraphics[angle=0,width=0.9 \textwidth]{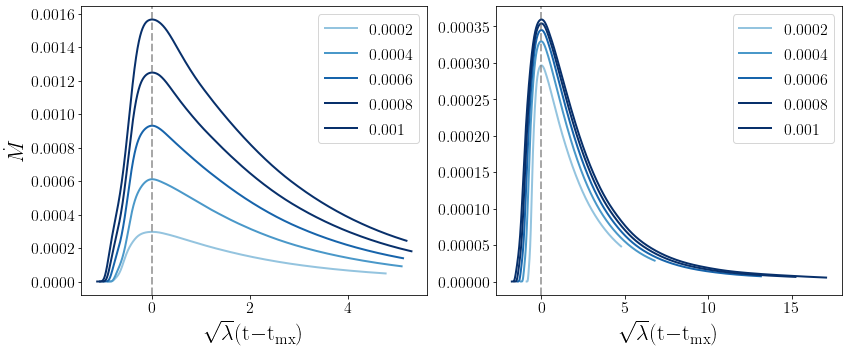}\\
\caption{Same as Fig.~\ref{fig:mdot1} but rescaling the axis by a factor $\sqrt{\lambda}$.}
\label{fig:mdot2}
\end{figure*}

Figures~\ref{fig:mdot1} and \ref{fig:mdot2} show the \bh{} mass growth rate as a function of time. In Fig.~\ref{fig:mdot1}, the left panel is for $\varphi_0 = 0.9$ cases, and the right panel is for  $\bar V_{,\varphi} = 1.458$ cases. Plots were created via Eq.~\ref{eq:mdot}. For the $\varphi_0 = 0.9$ cases in the left panel, the accretion rate increases with $\lambda$ since the steepness of the potential is determined by $\lambda$ and thus the larger value of $\varphi'$. In the right panel, we have the $\bar V_{,\varphi} = 1.458$ cases, namely the situations for which the scalar field initially experiences the same force. The main difference here is that $\dot M$ decays faster as one increases $\lambda$; the larger the value of $\lambda$, the steeper the potential, and the scalar field reaches the bottom of the potential faster. Figure~\ref{fig:mdot2} is the same as Fig.~\ref{fig:mdot1} but rescaling the time by a factor $\sqrt{\lambda}$ with the time-shifted such that the maxima of $\dot M$ are aligned. Since the rescaling factors out the $\lambda$ dependence, as expected, the cases with $\bar V_{,\varphi} = 1.458$ align with each other since the scalar field experiences the same initial force initially.

\begin{figure*}[!htbp]
\includegraphics[angle=0,width=0.9 \textwidth]{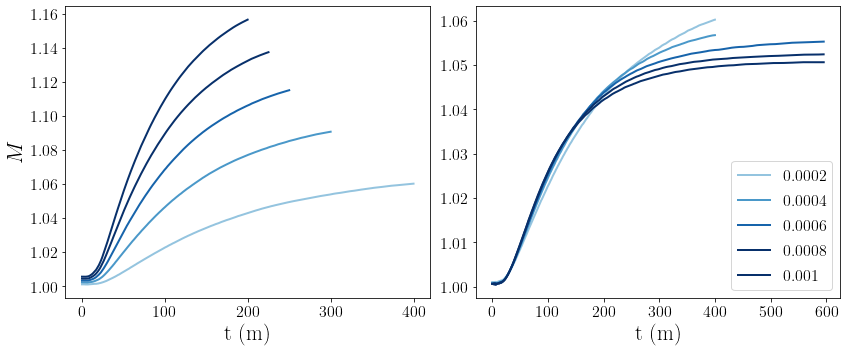}\\
\caption{\bh{} mass growth computed from the area of the apparent horizon as a function of time. Left panel is for the case $\varphi_0 = 0.9$, while the right panel is for the case $\bar V'(\varphi_0) = 5.832$.}
\label{fig:mass1}
\end{figure*}

\begin{figure*}[!htbp]
\includegraphics[angle=0,width=0.9\textwidth]{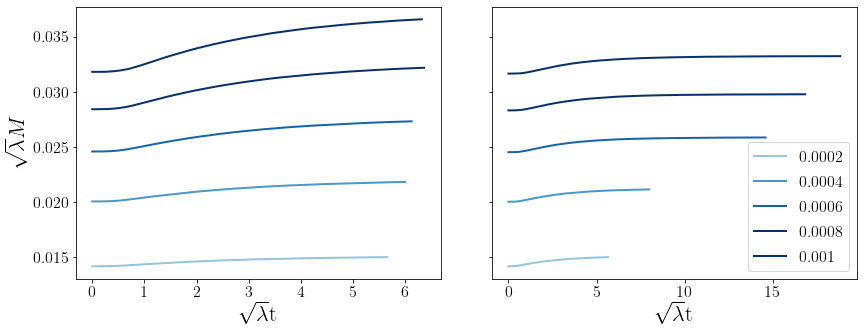}\\
\caption{Same as Fig.~\ref{fig:mass1} but rescaling the time by a factor $\sqrt{\lambda}$.}
\label{fig:mass2}
\end{figure*}

Figures~\ref{fig:mass1} and \ref{fig:mass2} show the \bh{} mass growth computed from the area of the apparent horizon as a function of time. In Figure~\ref{fig:mass1}, the left panel is for the case $\varphi_0 = 0.9$, while the right panel is for the case $\bar V_{,\varphi} = 5.832$. Figure~\ref{fig:mass2} is the same as Fig.~\ref{fig:mass1} but rescaling the axis by a factor $\sqrt{\lambda}$.

To gain a better understanding of the \bh{} mass growth formula as given by Eq.~(\ref{eq:mdot}), we will approximate
\begin{eqnarray}
     \varphi' & =& \mathring\varphi-\left(\frac{\beta^i}{\alpha}-s^i\right)\partial_i\varphi \approx \mathring\varphi 
\end{eqnarray}
since, as seen in Sec.~\ref{sec:scalar_dynamics}, the scalar field does not vary much across the computational domain. Therefore, Eq.~(\ref{eq:mdot}) becomes
\begin{eqnarray}
\dot M &\approx&\frac{8\,\sqrt{2}\,\pi\,\,M^2\,\dot\varphi^2}{1-32\,\pi\,M^2\,\left[\frac{1}{2}\dot\varphi^2+V\right]}\,,\label{eq:mdot2}
\end{eqnarray}
where we have also used $\Theta_{(k_\Sigma)} \approx - M^{-1}$ and $\alpha \approx 1$. In addition, our simulations fall under the slow-roll regime in which $\dot\varphi^2/2 \ll V$. Therefore
\begin{eqnarray}
\dot M &\approx&\frac{8\,\sqrt{2}\,\pi\,\,M^2\,\dot\varphi^2}{1-32\,\pi\,M^2\,V}\,.\label{eq:mdot3}
\end{eqnarray}
Lastly, our setup is such that $32\,\pi\,M^2\,V \ll 1$ and thus
\begin{eqnarray}
\dot M &\approx&  8\,\sqrt{2}\,\pi\,M^2\,\dot\varphi^2\,.\label{eq:mdot4}
\end{eqnarray}

Under the slow-roll approximation, Eqs.~(\ref{eq:friedmann}) and (\ref{eq:phi_friedmann}) take the form
\begin{eqnarray}
    H^2 &=& \frac{8\,\pi}{3}V\,,\label{eq:sr_friedmann}\\
    3H\dot{\varphi}&=&-V_{,\varphi} \,.\label{eq:sr_phi_friedmann}
\end{eqnarray}
With our potential $V = \lambda\,\varphi^4/4$, solutions to Eqs.~(\ref{eq:sr_friedmann}) and~(\ref{eq:sr_phi_friedmann}) yield 
\begin{equation}
\dot\varphi = -\sqrt{\frac{\lambda}{6\,\pi}}\varphi_*\exp{\left[-\sqrt{\frac{\lambda}{6\,\pi}}(t-t_*)\right]}\,,\label{eq:sl_phidot}
\end{equation}
where $t_*$ and $\varphi_*$ are the time and value of the scalar field when the system enters the slow-roll regime. Substitution of (\ref{eq:sl_phidot}) into Eq.~(\ref{eq:mdot4}) yields
\begin{equation}
\frac{1}{M} = \frac{1}{M_*}-\frac{1}{\widehat M}\left\lbrace1-\exp{\left[-\sqrt{\frac{2\,\lambda}{3\,\pi}}(t-t_*)\right]}\right\rbrace\,,\label{eq:sl_mbh}
\end{equation}
where
\begin{equation}
    \frac{1}{\widehat M} \equiv \sqrt{\frac{16\,\pi\,\lambda}{3}}\,\varphi^2_*\,.\label{eq:sl_mhat}
\end{equation}
For $t \gg t_*$, Eq.~(\ref{eq:sl_mbh}) becomes
\begin{equation}
\frac{M_*}{M} = 1-\frac{M_*}{\widehat M}\,.
\end{equation}
To grow a \bh{} with mass $M_*$ by a factor of $\xi = M/M_*$ would require 
\begin{equation}
    \xi = \left(1-\frac{M_*}{\widehat M}\right)^{-1}\,.
\end{equation}
For the values used in the present work of $\lambda \sim 10^{-4}m^{-2},\, \varphi_* \sim 1$, and $M_* \sim 1\,m$, one has that $M_*/\widehat M\sim 4\times 10^{-2}$, which yields $\xi \sim 1.04$, an estimate consistent with the results found in the previous section.
\\

\section{Conclusions}\label{sec:conclusions}

We have conducted a numerical relativity study of the accretion properties of a non-spinning black hole in a cosmology driven by a scalar field. Because the space-time is not asymptotically flat, we introduced modifications to the moving-puncture gauge condition to be able to function in cosmological space-times. We considered a scalar field with potential $ V=\lambda \,\varphi^4/4$ and varied the parameter $\lambda$ as well as the initial conditions for the scalar field. Using the dynamical horizon framework, we derived the black hole mass growth formula for these cosmological scenarios. We verified that the results from the simulations satisfy this mass growth formula. As with perturbative studies, we found that the accretion rate $\dot M \propto M^2$ with $M$ the mass of the black hole, and that $\dot M \propto \dot\varphi^2$. 
What was not anticipated was that the dynamics of the scalar field in the neighborhood of the black hole is not significantly different from the behavior of the field far away from the hole. We found situations in which the black hole can growth $\sim 15\%$ of its initial mass before the scalar field reaches the bottom of its potential. The next step is to investigate \bh{} scalar accretion for \bh{s} with spin and linear momentum.

\section{Acknowledgments}

This work is supported by NSF grants PHY-2411068 and PHY-2207780. 


\end{document}